\documentclass[a4paper, 12pt]{article}
\usepackage[english]{babel}
\usepackage[cp1250]{inputenc}
\usepackage[T1]{fontenc}
\usepackage{epsfig}
\usepackage{latexsym}
\newcommand{\un}{\underline}

\begin{document}
\title{Chain of impacting pendulums as non-analytically perturbed sine-Gordon system
\thanks{Paper supported in part by ESF ``COSLAB'' Programme  }}
\author{H. Arod\'z  and  P. Klimas \\
Marian Smoluchowski Institute of Physics, Jagiellonian University, \\
Reymonta 4, 30-059 Cracow, Poland}
\date{$\;\;\;$}
\maketitle

\vspace*{2cm}

\begin{abstract}
We investigate a mechanical system consisting of infinite number of harmonically coupled pendulums which can
impact on two rigid rods. Because of gravitational force the system has two degenerate ground states. The
related topological kink -- likely the simplest one presented in literature so far -- is a compacton, that is it
has strictly finite extension. In the present paper we elucidate the relation of such system with sine-Gordon
model. Also, solutions describing waves with large amplitude, and an asymptotic formula for the width of the
kink are obtained.
\end{abstract}

\vspace*{2cm}
\noindent


\pagebreak

\section{Introduction}

Theoretical studies of classical mechanical systems with large number of degrees of freedom are relevant to many
branches of condensed matter physics (not to mention engineering sciences), where quantal aspects of microscopic
world are practically not visible. Moreover, since the seminal work of Fermi, Pasta and Ulam \cite{1} it is
clear that such systems can also provide useful models on which concepts of theoretical physics can be tested or
illustrated. For example, studies of anharmonic lattices give insights into nonlinear transport phenomena, see
e.g., \cite{2}, or mechanisms of thermal conductivity, see, e.g., \cite{3}. Another example: sine-Gordon
solitons, which appear in many places including quantum field theory (the duality with Thirring model), have
beautiful realisation in a system composed of harmonically coupled pendulums, see, e.g., \cite{4}.

In recent papers \cite{5, 6} we have investigated a mechanical system  -- described below in Section 2 -- which
has infinite number of degrees of freedom. It exhibits spontaneous symmetry breaking and  related to it
topological kink and antikink. Its main attractive feature is extraordinary simplicity of the kink. It has
strictly finite extension equal to $\pi$ in a certain limit, and in that limit the kink is represented by the
piece of sine function in the interval $[ - \pi/2, \: \pi/2].$ To the best of our knowledge, it is the simplest
topological defect presented in literature so far. Furthermore, time evolution of the system during the symmetry
breaking transition can be described analytically. For these reasons, the system is quite interesting as a
testing ground for various ideas of theory of topological defects. Physics of such objects poses many very
interesting albeit hard to solve dynamical questions, see \cite{7, 8} for overviews.

The purpose of the present paper is to supplement our previous investigations \cite{5, 6}. In next Section we
briefly recall the model and certain results in order to make the paper self-contained. We also introduce a
folding transformation which maps trajectories of certain auxilliary, simpler model into trajectories of our
system. Section 3 is devoted to large amplitude waves -- the ones having small amplitudes were investigated in
\cite{6}. The two types of waves involve very different kinds of motion.  In Section 4 we calculate the width of
the kink in a limit when it increases to infinity. Connection of our system with the one considered in
connection with the sine-Gordon soliton is elucidated in Section 5. We show that from mathematical viewpoint the
unfolded version of our model can be regarded as non-analytically perturbed sine-Gordon system. In particular,
the sine-Gordon soliton and breather provide approximate solutions of our model. Nevertheless, our system and
the sine-Gordon system are very different from physical viewpoint. Finally, in Section 6 we point out several
dynamical problems which can be posed within our model and which, in our opinion, deserve a careful
investigations.
\\

\section{The model and the folding transformation }

We consider one dimensional model with the  Lagrangian
\begin{equation}
L= \frac{1}{2} (\partial_{\tau}\phi)^2 - \frac{1}{2} (\partial_{\xi}\phi)^2
- V(\phi),
\end{equation}
where  the field potential $V(\phi)$ has the form
\begin{equation}
V(\phi) = \left\{
\begin{array}{lcl}
\cos\phi -1  &  {\rm for} &  | \phi| \leq \phi_0 \\
\infty  &  {\rm for}  &   |\phi| > \phi_0,
\end{array}
\right.
\end{equation}
see Fig.1. Here  $\phi_0$ is a constant, $\pi >\phi_0 >0$. Thus, all values of the real field $\phi$ are
restricted to the interval $[- \phi_0, \: \phi_0].$ The potential has two symmetric minima at $\phi= \pm
\phi_0$, and it is not continuous at them.
\begin{center}
\hspace*{0.5cm}  \epsfig{file=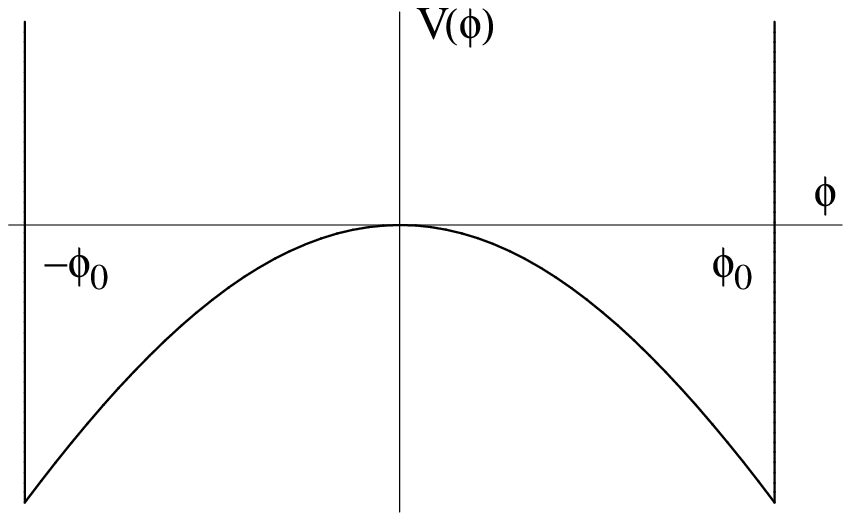}
\end{center}
\begin{center}
Fig.1. The potential $V(\phi)$
\end{center}
This model describes an infinite set of harmonically coupled pendulums in a homogeneous gravitational field:
$\xi$ is a dimensionless coordinate along a wire to which the pendulums are attached, $ \tau$ is a dimensionless
time variable, and $\phi(\xi, \tau)$ is equal to the angle between the upward vertical direction and the arm of
the pendulum located at the point $\xi$ and time $\tau$. The pendulums can move only in planes perpendicular to
the wire. The restriction
\begin{equation}
-\phi_0 \leq \phi \leq \phi_0
\end{equation}
is enforced by two rigid rods parallel to the wire. We assume that the pendulums bounce elastically from the
rods. For a more detailed description of this system see \cite{5}.

Time evolution of $ \phi(\xi, \tau)$ is governed by the following equation
\begin{equation}
(\partial_{\tau}^2 - \partial_{\xi}^2)\phi = \sin \phi \;\;\; \mbox{if} \;\;\; |\phi| < \phi_0,
\end{equation}
and by the elastic bouncing condition
\begin{equation}
\partial_{\tau}\phi(\xi, \tau) \rightarrow - \partial_{\tau}\phi(\xi, \tau) \;\;\; \mbox{if} \;\;\; \phi(\xi, \tau) = \pm
\phi_0.
\end{equation}
Hence, the time evolution of the velocities $\partial_{\tau}\phi$  is in general not continuous.
 The vertical upward position, that is $\phi =0$, is unstable. The stable positions are given by $\phi =
\pm \phi_0.$ Energy of the system is conserved. Typical motion of a pendulum consists of smooth `flights'
bewteen impacts with the rigid rods.

In the case $ \phi_0 \ll 1$ we may replace $\sin \phi $ by $\phi$, and istead of Eq.(4) we then have
\begin{equation}
(\partial_{\tau}^2 - \partial_{\xi}^2)\phi =  \phi \;\;\; \mbox{if} \;\;\; |\phi| < \phi_0.
\end{equation}

Physically, the kink is obtained when the pendulums gradually turn from one rod to the other one along the wire.
Its center is identified with the point $\xi_1$ such that $ \phi(\xi_1) = 0$ -- the corresponding pendulum
points in the upward direction. Static kink is represented by a time-independent solution $\phi_c(\xi)$ of Eq.
(4) or Eq.(6) which interpolates between $ - \phi_0$ and $+\phi_0$, and such that $ \phi_c$ and $
\partial_{\xi}\phi_c$ are continuous when $ \phi_c \rightarrow \pm \phi_0.$ Such solutions have particularly
simple form in the case of Eq. (6): for example, the kink located at $ \xi_1 =0$ is given just  by
\begin{equation}
\phi_c(\xi) = \left\{\begin{array}{ccc} - \phi_0 & \mbox{for} & \xi \leq - \frac{\pi}{2}, \\
\phi_0 \sin\xi &  \mbox{for} & - \frac{\pi}{2} \leq  \xi \leq  \frac{\pi}{2},  \\
\phi_0 & \mbox{for} & \xi \geq  \frac{\pi}{2}.  \end{array} \right.
\end{equation}
Anti-kink is represented by $ \phi_{\overline{c}}(\xi) = - \phi_c(\xi).$ Translations or `Lorentz' boosts give,
respectively, shifted or moving kink or anti-kink. As discussed in \cite{5}, the kink has strictly finite
extension for any $ \phi_0 < \pi.$ Therefore, it is a simple example of so called compactons which were found
for modified Korteweg-deVries models \cite{9, 10, 11}, and in a system of pendulums with non-trivial anharmonic
coupling between them \cite{12}.

Condition (5) is quite troublesome because it introduces discontinuity in phase space trajectories of pendulums.
For this reason, often it is convenient to `unfold' the model. By this we mean passing to a new model with a new
field $\un{\phi}(\xi, \tau)$ such that $\partial_{\tau}\un{\phi}$ has continuous time evolution. $ \un{\phi}$
can take arbitrary real values. The relation between $\phi$ and $ \un{\phi}$ has the following form. If
\begin{equation}
\un{\phi} \in [-\phi_0 + 2 k \phi_0, \phi_0 + 2 k \phi_0],
\end{equation}
where $k$ is an integer, then
\begin{equation}
\phi(\xi, \tau) = \left\{
\begin{array}{lcl}
\underline{\phi}(\xi, \tau) - 2 k \phi_0 & \mbox{if} &    k \;\; \mbox{is even} \\
2 k \phi_0 - \underline{\phi}(\xi, \tau) & \mbox{if} &  k \;\; \mbox{is odd}.
\end{array}\right.
\end{equation}
The impacts on the rods occur when $\un{\phi} = \phi_0 + 2 l \phi_0$ with integer $l$. Relation (9)  is depicted
in Fig. 2.
\begin{center}
\hspace*{-2cm}\epsfig{file=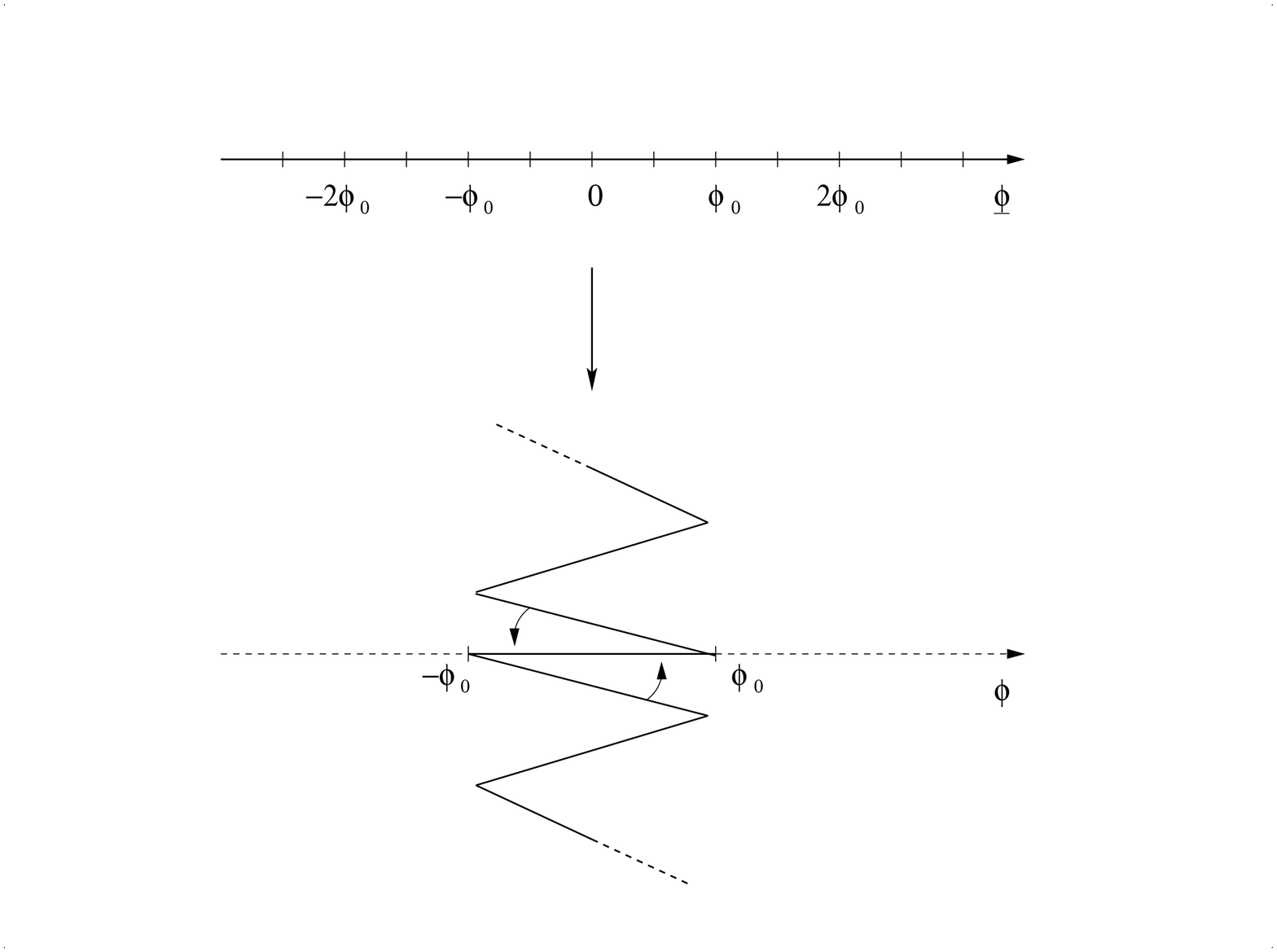,height=10cm,width=18cm}
\end{center}
\begin{center}
Fig.2. The folding transformation
\end{center}
\vspace*{0.5cm}

Evolution of $\underline{\phi}$ is governed by the equation
\begin{equation}
(\partial_{\tau}^2 - \partial_{\xi}^2)\underline{\phi} = - \frac{d \un{V}(\un{\phi})}{ d \un{\phi}},
\end{equation}
where $ \un{V}(\un{\phi})$ is given by the following formula:  if $ \un{\phi}$ belongs to the interval (8), then
in the case of Eq.(4)
\[
\underline{V}(\underline{\phi}) = \cos(\un{\phi} - 2 k \phi_0) -1,
\]
see Fig. 3,
\begin{center}
\hspace*{-2cm} \epsfig{file=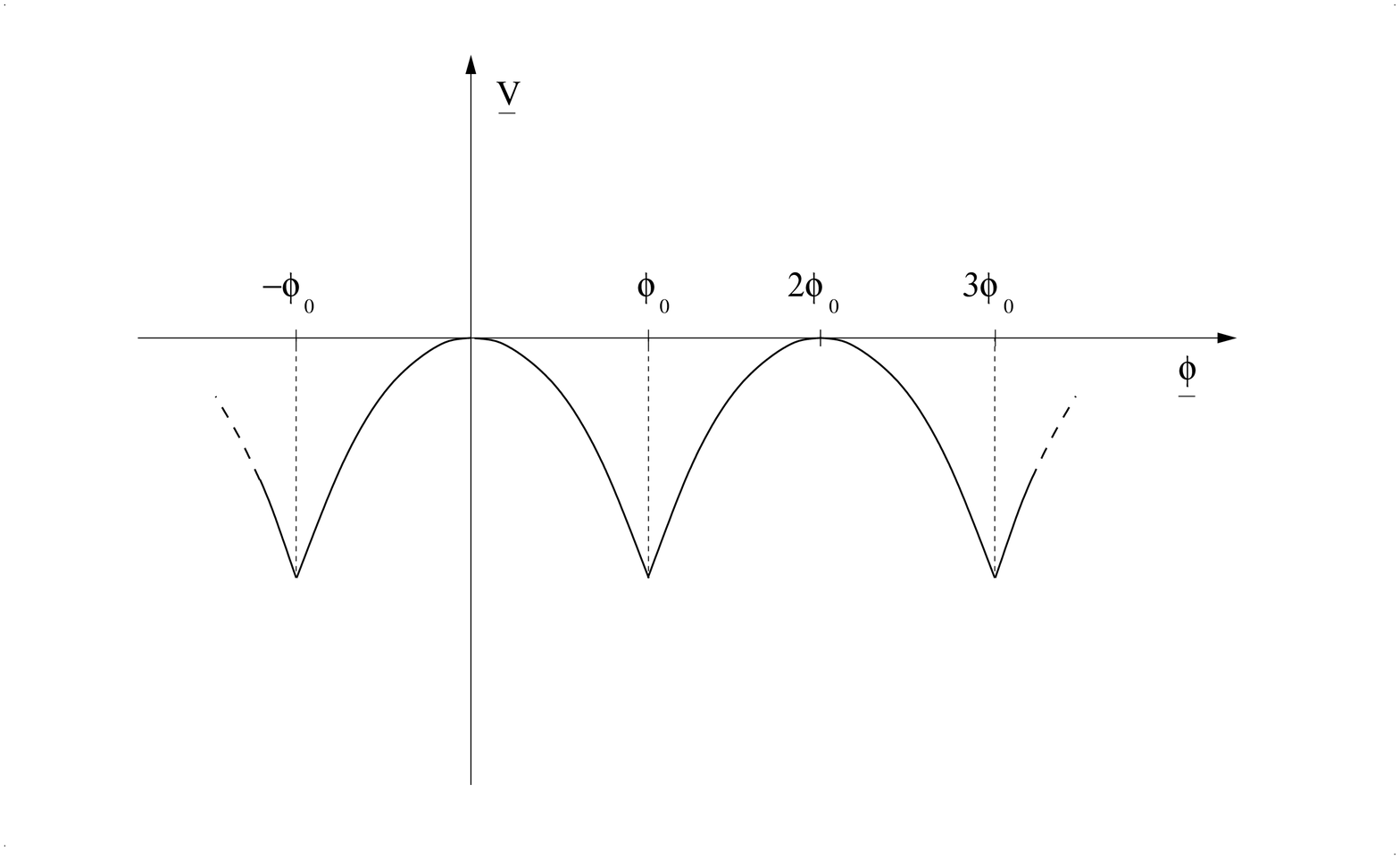,height=8cm,width=16cm}
\end{center}
\begin{center}
Fig.3. The potential $\underline{V}(\underline{\phi})$
\end{center}
\vspace*{0.5cm} \noindent and
\begin{equation}
\underline{V}(\underline{\phi}) = - \frac{1}{2} (\un{\phi} - 2 k \phi_0)^2
\end{equation}
in the case of Eq.(6).

Time evolution of $\underline{\phi}$ and of its first derivatives $\partial_{\tau}\underline{\phi}, \;\;
\partial_{\xi}\underline{\phi}$ is continuous. It follows from transformation (9) that condition (5) is
automatically satisfied. \\

\section{Waves of impacting pendulums }

Let us first consider uniform motions of the pendulums, i.e., the ones such that $\phi$ does not depend on
$\xi$. In this case Eq.(6) is reduced to
\begin{equation}
\frac{d^2 \phi}{ d \tau^2} = \phi
\end{equation}
if
\[
| \phi | < \phi_0.
\]
The energy integral of motion corresponding  to Eq.(12) has the form
\begin{equation}
\dot{\phi}^2 - \phi^2 = c_0.
\end{equation}
In the case the constant $c_0$ is positive we may write $c_0 = u^2,$ where $u > 0$ is the absolute value of the
velocity of pendulums at the vertical upward position ($\phi =0$). This class of trajectories describes the
pendulums impacting simultaneously on one barrier rod with the velocity $ (u^2 + \phi_0^2)^{1/2}$, elastically
bouncing back from it and moving over the upward positon and further, until they impact on the other rod, from
which they are reflected again -- the motion is periodic in time. The pertinent solutions of Eq.(12) have the
form
\begin{equation}
\phi(\tau) = \pm u \sinh(\tau - \tau_0).
\end{equation}
At the time $\tau_0$ all pendulums are in the vertical position. They fall onto one of the rods at the time
$\tau_q$ such that
\begin{equation}
u \sinh(\tau_q - \tau_0) = \phi_0.
\end{equation}
Hence,
\begin{equation}
\tau_q - \tau_0 = \mbox{Arsinh}(\phi_0/u) = \ln\left(\frac{\phi_0}{u} + \sqrt{ 1 + \frac{\phi_0^2}{u^2}}\right).
\end{equation}
The period $T_0$ of this motion is given by the formula
\begin{equation}
T_0 = 4 (\tau_q - \tau_0).
\end{equation}
The phase space picture of trajectories of this type is presented in Fig. 4 by the curve $A$. Because the
velocity changes its sign at the moment of the impact, the trajectory consist of two disconnected parts.

The curves $B$ and $C$ in Fig. 4 represent two different sets of trajectories such that the pendulums do not
pass through the vertical position ($ \phi =0$): the pendulums bounce from one rod, raise a little bit and fall
on the same rod. Also this motion is periodic in time. The corresponding solutions of Eq.(12) have the form
\begin{equation}
\phi(\tau) = \pm \phi_m \cosh(\tau - \tau_0),
\end{equation}
where $\phi_m$ is a constant such that $ 0 < \phi_m < \phi_0.$ In this case, $c_0 = - \phi_m^2$ in formula (13).
At the time $\tau_0$ pendulums reach their highest position, $\phi(\tau_0) = \pm \phi_m.$ They fall on the rod
after the time interval $\tau_q - \tau_0,$ where
\begin{equation}
\cosh(\tau_q - \tau_0) = \frac{\phi_0}{\phi_m}.
\end{equation}
Hence,
\begin{equation}
\tau_q - \tau_0 = \ln\left(\frac{\phi_0}{\phi_m} + \sqrt{ \frac{\phi_0^2}{\phi_m^2} - 1}\right).
\end{equation}
The period is equal to $T_0 = 4(\tau_q - \tau_0)$. Waves generated from these solutions were discussed in
\cite{5}.

\begin{center}
\epsfig{file=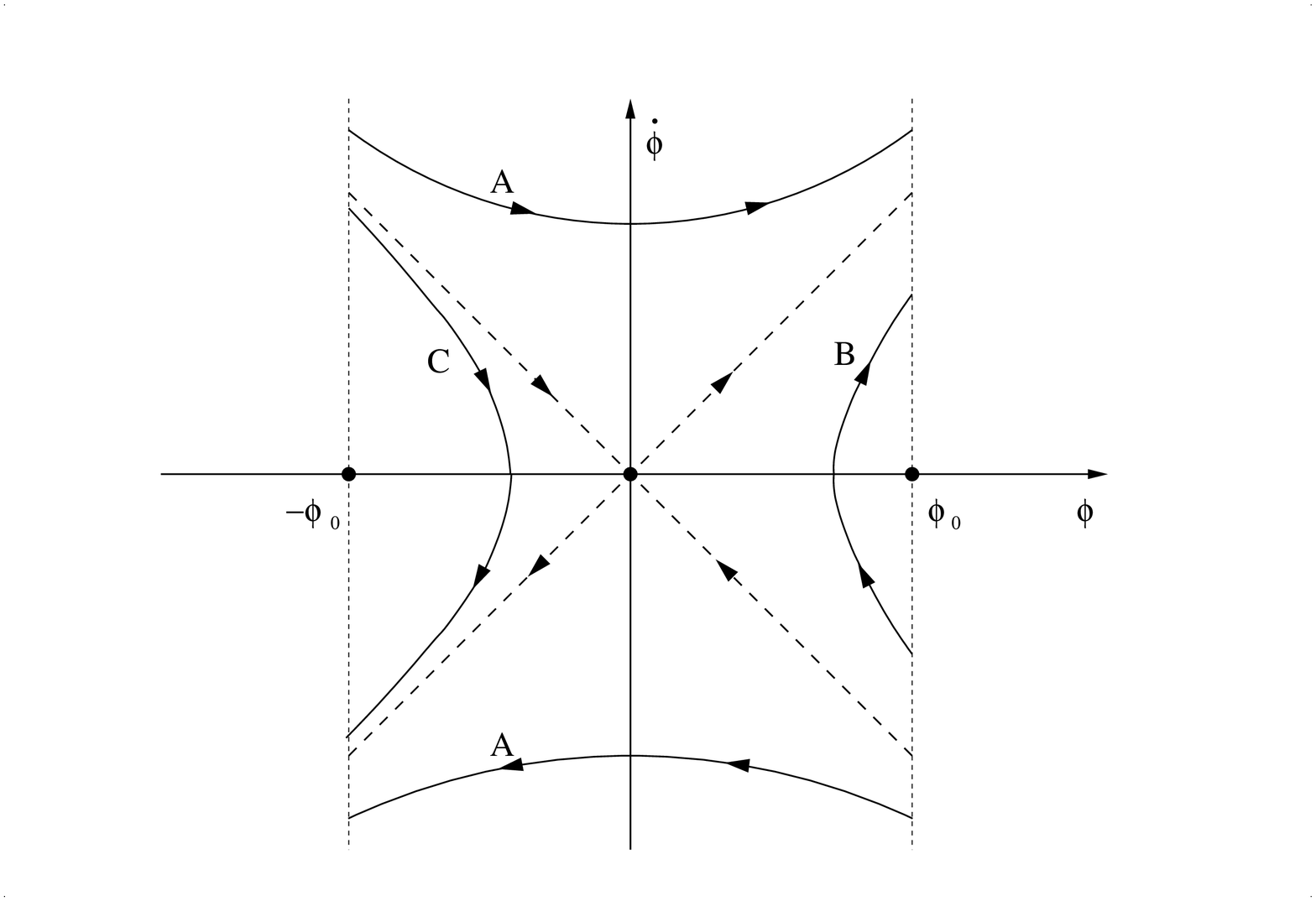,height=6cm,width=14cm}
\end{center}
\begin{center}
Fig.4. Phase space trajectories of the uniformly impacting pendulums
\end{center}

There are solutions of Eq.(12) for which $c_0 = 0$:
\begin{equation}
\phi_{\pm}(\tau) = A \phi_0 \exp[ \pm (\tau - \tau_0)],
\end{equation}
where $A= \pm 1.$ The solution $\phi_+$ describes pendulums which at the time  $\tau_0$ start to move up  from
one of the rods with the initial velocity $A \phi_0$ and reach the upward position at $ \tau = + \infty.$ The
other solution describes the reverse process. These motions are not periodic ($T_0 = \infty$), and they separate
the previous three classes of trajectories. In Fig. 4 they are represented by the dashed lines.

Finally, the thick dot in the center of Fig. 4 denotes the trivial, unstable trajectory $\phi=0, $ and the thick
dots at $\phi = \pm \phi_0$ the stable equilibrium positions.

The spatially homogeneous solutions discussed above can easily be generalized to solutions describing periodic
in time and space waves just by performing a Lorentz-like boost. For example, in the familiar case of
Klein-Gordon wave equation for a scalar field $\psi$, $\Box \psi = m_0^2 \psi,$ instead of Eq.(12) we have  $
d^2 \psi/ d \tau^2 = - m_0^2 \psi$ with the solutions $ \psi = A \exp(\pm i m_0 \tau).$ Lorentz boost with
velocity $w$, where $|w| < 1$, amounts to the substitution $ \tau \rightarrow \zeta = \gamma (\tau - w \xi),$
where $ \gamma = (1 - w^2)^{-1/2}.$ In this way we obtain the plane wave with the frequency $ \omega = \pm
\gamma m_0$ and the wave vector $k = \pm m_0 \gamma w$ such that $ \omega^2 - k^2 = m_0^2.$

Applying the boost to the solutions of the class $B$ or $C$ we obtain waves which  coincide with the waves
discussed in our previous work \cite{6} after the substitution $w = 1/v$. Waves obtained from solution $A$ are
combined piecewise from the functions
\begin{equation}
\phi(\tau, \xi) = \pm u \sinh(\zeta - \tau_0),
\end{equation}
see Fig. 5.
\begin{center}
\epsfig{file=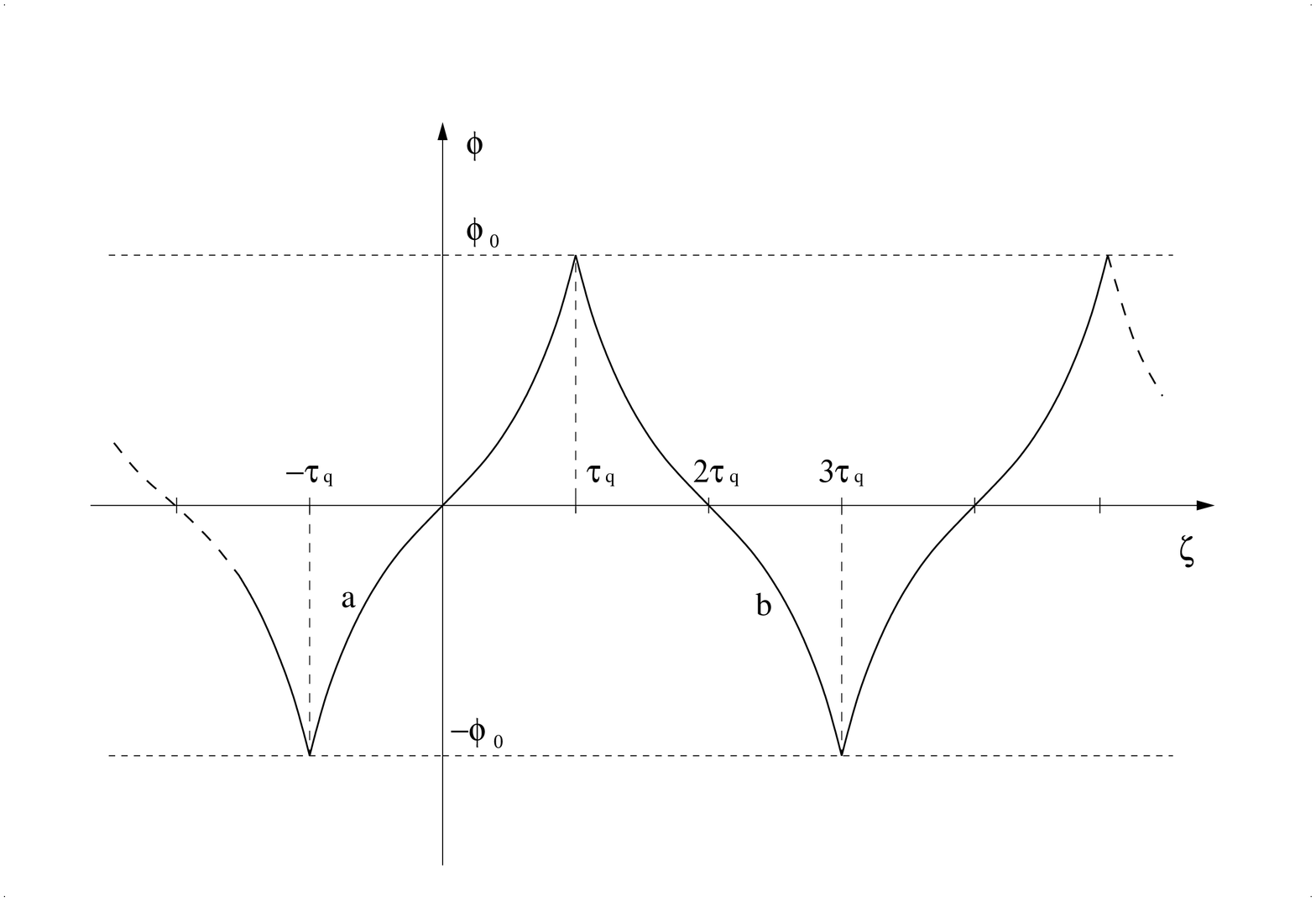,height=9cm,width=13cm}
\end{center}
\begin{center}
Fig.5. The wave of impacting pendulums
\end{center}
Here
\begin{equation}
\zeta = \frac{v \tau - \xi}{\sqrt{v^2 -1}},
\end{equation}
$\tau_0$ is a constant, and $v$ is the phase velocity of the wave. It is assumed that $|v| >1.$ The group
velocity is equal to the boost velocity $w = 1/v.$ The wave length and the frequency are equal to
\[
\lambda_0 = 4 \sqrt{v^2 -1} ( \tau_q - \tau_0), \;\;\; \omega_0 = \frac{v}{\sqrt{v^2 -1}} \frac{1}{T_0},
\]
where $T_0$ is given by formula (17). Depending on the value of $v$, $\lambda_0$ can be any positive number,
while $0 < |\omega_0| < 1/T_0.$ When $ v \rightarrow \infty$ we recover the uniformly bouncing pendulums: $
\lambda_0 = \infty, \;\; \omega_0 = T_0^{-1}.$

The trick with boost can be applied if we have a solution which exists for all $\tau$ -- otherwise we would
obtain $\phi(\xi, \tau)$ on a part of the $\xi$ axis only. For this reason, solutions (21) have to be combined
together in order to cover the whole time axis. There are two possibilities which differ by overall sign factor:
\[
\phi(\tau) = \left\{ \begin{array}{ccc} \phi_0 \exp(\tau - \tau_0) & \mbox{for} & \tau \leq \tau_0, \\
\phi_0 \exp(\tau_0 - \tau) & \mbox{for} & \tau \geq \tau_0,
\end{array} \right.
\]
and $ - \phi(\tau).$ Boosting these solutions consists of replacing $\tau$ by $ \zeta$ as above. We obtain a
travelling wave in which the pendulum located at $\xi_1$ hits the rod at the  time
\[ \tau_1 = \frac{\sqrt{v^2-1}}{v}\: \tau_0 + \frac{\xi_1}{v},
\]
while all other pendulums either are raising (those at $\xi < \xi_1$) or falling down (those at $ \xi > \xi_1$).
\\

\section{The width of the compacton when $\phi_0 \rightarrow \pi- $}

In the particular case when $\phi_0 = \pi$ the two rods merge into one put below the wire which supports the
pendulums, and the angle $\phi$ can take arbitrary value in the interval $[-\phi_0, \phi_0].$ As far as the
static configurations are considered, such a system of pendulums coincides with the one corresponding to the
sine-Gordon equation, where the rod is absent altogether. The point is that the least energy configurations of
the systems coincide. Of course, for time-dependent configurations the two systems remain different because in
our system the pendulums bounce from the rod, while in the sine-Gordon one they can move without encountering
any obstacles.

The soliton of sine-Gordon model has infinite extension because of its exponential tails.  On the other hand,
the compacton does not have tails for any $ \phi_0 < \pi.$ Therefore, its width $\xi_0(\phi_0)$ should be
divergent when $ \phi_0 \rightarrow \pi$ from below. In \cite{5} we have obtained the following formula for the
width
\begin{equation}
\xi_0(\phi_0) = \frac{\phi_0}{\sqrt{2}} \int^1_0 d\lambda \frac{1}{\sqrt{\cos(\lambda \phi_0) - \cos\phi_0}}.
\end{equation}
Substituting here $ \phi_0 = \pi - \epsilon, \: \epsilon >0 $ and changing the integration variable to $ \sigma
= (\pi - \epsilon)\lambda / 2$ we obtain
\begin{equation}
\xi_0(\phi_0) =  \int^{\pi/2 - \epsilon/2}_0 d\sigma \frac{1}{\sqrt{\cos^2(\epsilon/2) - \sin^2\sigma}}.
\end{equation}
Second change of integration variable, namely
\[
\sigma \rightarrow s = \frac{\sin\sigma}{\cos(\epsilon/2)}
\]
yields the elliptic integral of the first kind
\begin{equation}
\xi_0(\phi_0) = \int^1_0 d s \frac{1}{\sqrt{(1 - s^2 \cos^2(\epsilon/2)) ( 1 - s^2)}}.
\end{equation}
Thus,
\begin{equation}
\xi_0(\pi - \epsilon) = K(\cos^2 \frac{\epsilon}{2}).
\end{equation}
Because
\[
K(x) \cong \frac{1}{2} \ln \frac{16}{1 - x}
\]
for $x \rightarrow 1-,$ see e.g. \cite{13}, we finally obtain that
\begin{equation}
\xi_0(\pi - \epsilon) \cong \ln\frac{4}{\sin\frac{\epsilon}{2}}
\end{equation}
when $ \epsilon \rightarrow 0+.$  Thus, the width of the compacton diverges logarithmically
when $\phi_0 \rightarrow \pi-$.\\

\section{Equivalence with the non-analytically perturbed sine-Gordon system}

The potential $\underline{V}(\underline{\phi})$ is periodic in $ \un{\phi}$, hence it can be written as Fourier
series. Straightforward calculations show that potential (11) can be written in the form
\begin{equation}
\un{V}(\un{\phi}) = - \frac{\phi_0^2}{6} \left[ 1 + \frac{12}{\pi^2} \sum^{\infty}_{n=1} \frac{(-1)^n}{n^2}
\cos(\frac{n \pi}{\phi_0} \un{\phi})\right].
\end{equation}
It is clear  from Fig. 3 that the derivative  $ \un{V} $ with respect to $ \un{\phi}$ does not exist at the
points $ \un{\phi} = k \phi_0$, where $ k$ is an odd integer. At these points the Fourier series for $ d
\un{V}(\un{\phi}) / d \un{\phi}$  is divergent.

The constant and the $n=1$ terms in formula (29) give the potential
\begin{equation}
\un{V}^{(1)}(\un{\phi}) = - \frac{\phi_0^2}{6} + \frac{2\phi_0^2}{\pi^2}\cos(\frac{\pi}{\phi_0} \un{\phi}),
\end{equation}
which is analytic in $ \un{\phi}$. Let us introduce the new field
\begin{equation}
\psi(\sqrt{2}\tau, \sqrt{2}\xi) = \frac{\pi}{\phi_0} \un{\phi}(\xi, \tau) + \pi.
\end{equation}
Euler - Lagrange equation obtained from the potential $\un{V}^{(1)}(\un{\phi})$ can be written in the form
\begin{equation}
(\partial^2_{\tau} - \partial^2_{\xi}) \psi(\tau, \xi) = - \sin \psi(\tau, \xi),
\end{equation}
which is identical with the sine-Gordon equation (on both sides we have replaced $\sqrt{2} \tau, \sqrt{2}\xi$ by
$\tau, \xi$).

The $ n > 1$ terms in formula (29) give the non-analytic perturbation of the sine-Gordon potential (30),
\[
\un{V}^{pert}(\psi) = - \sum^{\infty}_{n=2} \frac{1}{n^2} \cos(n \psi),
\]
and of the sine-Gordon equation
\[
(\partial^2_{\tau} - \partial^2_{\xi}) \psi = - \sin \psi - \sum^{\infty}_{n=2} \frac{ \sin(n \psi)}{n}.
\]
The sum on the r.h.s. of the last equation is finite for all values of $\psi$, but it is not continuous at $\psi
= 2 k \pi,$ where $k$ is integer. This follows from the formula
\[
\sum^{\infty}_{n=2} \frac{ \sin(n \psi)}{n} = \frac{i}{2} \ln(- e^{i \psi}) - \sin\psi
\]
and the fact that $\ln$ function has a cut along the negative part of the real axis.

The replacement of $\un{V}(\un{\phi})$ by $\un{V}^{(1)}(\un{\phi})$ gives the sine-Gordon approximation to our
system. The folding transformation applied to $\un{V}^{(1)}(\un{\phi})$ yields the potential
\[
V^{(1)}(\phi) = - \frac{\phi_0^2}{6} + \frac{2\phi_0^2}{\pi^2}\cos(\frac{\pi}{\phi_0} \phi)
\]
where $ - \phi_0 \leq \phi \leq \phi_0.$ It is  presented in Fig. 6. The vertical lines at $ \phi = \pm \phi_0$
in Fig. 6 correspond to the two rods. The main difference between this potential and the original one, see Fig.
1, is that now the first derivative of the potential vanishes at each of the two minima. This fact has profound
influence on the shape  of the soliton at large $\xi$: the soliton acquires the exponential tails.

\begin{center}
  \epsfig{file=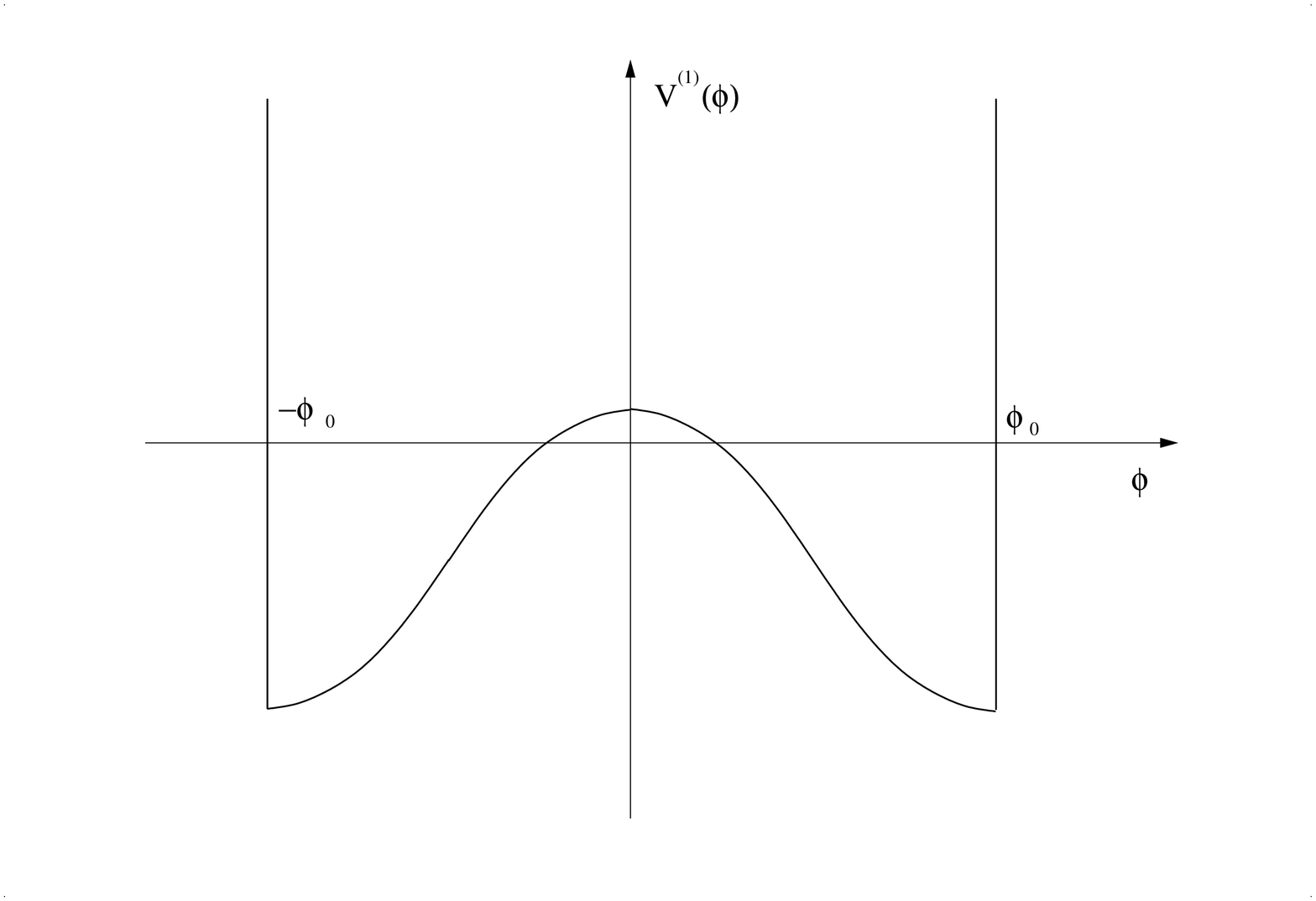,height=9cm,width=14cm}
\end{center}
\begin{center}
Fig.6. The potential $V^{(1)}(\phi).$
\end{center}
\vspace*{0.5cm}

It is clear that with the help of the sine-Gordon approximation and formula (9) we can utilise the well-known
exact solutions of sine-Gordon equation in order to obtain approximate solutions of our system. For example, the
sine-Gordon breather \cite{4} gives
\begin{equation}
\un{\phi}(\xi, \tau) = \frac{4 \phi_0}{\pi} \arctan\left( \frac{\sin\left(\frac{\sqrt{2}\: p\:
\tau}{\sqrt{1+p^2}}\right) }{p \:\cosh\left(\frac{\sqrt{2}\: \xi}{\sqrt{1+p^2}}\right)}\right) - \phi_0,
\end{equation}
where $ 0 < p < \infty$ is a parameter.  The folding transformation transforms this solution into the
approximate trajectory of our system of pendulums: at the time $ \tau =0$ all pendulums are just bouncing from
the rod at $ \phi = - \phi_0$, next they move up. Some of them  can cross the upward vertical position provided
that  $ p \leq 1,$ but they do not reach the other rod. All pendulums stop simultaneously at the time
\[
\tau_m = \frac{\pi \sqrt{1 + p^2}}{2 \sqrt{2} p},
\]
and they begin to slide back to the rod, from which they bounce again at the time $2 \tau_m.$. The motion is
periodic in time. We do not know whether there exists breather in our original system with evolution equation
(4) or (6).
\\

\section{Remarks}

\noindent 1. Our model has several attractive features: simplicity of differential equation (6); the simple
analytical form (7) of the kink; neat physical realisation as the system of pendulums with their motions
restricted by the two rigid rods. For these reasons we believe that the model can be useful for case studies of
such interesting yet difficult to understand in detail phenomena as production of topological defects during
symmetry breaking transitions (certain preliminary results in this direction were presented in \cite{6}),
scattering of kink on anti-kink, or
interaction of kink with a boundary in analogy to investigations reported in \cite{14}. \\

\noindent 2. Mechanical systems with impact have been investigated recently in connection with so called grazing
bifurcation as a road to chaotic behaviour, see, e. g., \cite{15, 16} -- the dynamics of such systems is far
from trivial. Our system is of field theoretical type, that is it has infinite number of degrees of freedom,
while most studies presented in literature so far are devoted to finite systems. It would be very interesting to
find out whether the chaotic motions are present in our system and, if present, what is their role in dynamics
of
compactons. \\

\end{document}